\documentclass[galaxies,article,accept,moreauthors,pdftex]{Definitions/mdpi} 

\newcommand{\be}{\begin{equation}}
\newcommand{\ee}{\end{equation}}
\newcommand{\bea}{\begin{eqnarray}}
\newcommand{\eea}{\end{eqnarray}}

\newcommand{\mnras}{Mon.~Not. R. Astron.~Soc.}
\newcommand{\apj}{Astrophys. J.}
\newcommand{\aap}{Astron. Astrophys.}

\newcommand{\pasp}{Publ. Astron. Soc. Pac.}

\newcommand{\apjs}{Astrophys. J. Suppl.}

\newcommand{\apjl}{Astrophys. J. Lett.}

\newcommand{\aapr}{Astrophys. Space Phys. Rev.}
\newcommand{\pasj}{Publications of the Astronomical Society of Japan}

\firstpage{1} 
\pubvolume{1}
\issuenum{1}
\articlenumber{0}
\pubyear{2021}
\copyrightyear{2020}
\externaleditor{Academic Editor: Firstname Lastname} 
\datereceived{} 
\dateaccepted{} 
\datepublished{} 
\hreflink{https://doi.org/} 

\Title{\boldmath{$\gamma$}-ray Flux and Spectral Variability of Blazar Ton 599 during Its 2021 Flare}

\TitleCitation{{$\gamma$}-ray Flux and Spectral Variability of Blazar Ton 599 during Its 2021 Flare}

\Author{Bhoomika 
 Rajput * and Ashwani Pandey \orcidB{}{}}
\AuthorCitation{Rajput, B.; Pandey, A.}
\address[1]{Indian Institute of Astrophysics, Block II Koramangala, Bangalore 560034, India;  ashwanitapan@gmail.com}

\corres{\hangafter=1 \hangindent=1.05em \hspace{-0.82em} Correspondence: bhoomikarjpt2@gmail.com}

\abstract{Blazars are known to emit exceptionally variable non-thermal emission over the wide range (from radio to $\gamma$-rays) of electromagnetic spectrum. We present here the results of our $\gamma$-ray flux and spectral variability study of the blazar Ton 599, which has been recently observed in the $\gamma$-ray flaring state. Using 0.1$-$300 GeV $\gamma$-ray data from the $\it{Fermi}$ Gamma-ray Space Telescope (hereinafter $\it{Fermi}$), we generated one-day binned light curve of Ton 599 for a period of about one-year from MJD 59,093 to MJD 59,457. During this one year period, the maximum $\gamma$-ray flux detected was \mbox{2.24 $\pm$ 0.25 $\times$ $10^{-6}$ ph cm$^{-2}$ s$^{-1}$ at MJD 59,399.50}.  We identified three different flux states, namely, epoch A (quiescent), epoch B (pre-flare) and epoch C (main-flare). For each epoch, we {calculated} the $\gamma$-ray flux variability amplitude (F$_{var}$) and found that the source showed largest flux variations in epoch C with F$_{var} \sim$ 35\%. We modelled the $\gamma$-ray spectra for each epoch and found that the Log-parabola model adequately describes the $\gamma$-ray spectra for all the three epochs. {We estimated the size of the $\gamma$-ray emitting region as 1.03 $\times$ $10^{16}$ cm and determined that the origin of $\gamma$-ray radiation, during the main-flare, could be outside of the broad line region.}}


\begin{document}


\section{Introduction} \label{sec:intro}
Blazars are the jetted subclass of the Active Galactic Nuclei (AGNs) that are understood as AGNs with very small viewing angles to the line of sight \citep{1995PASP..107..803U, 2017A&ARv..25....2P}. Blazars emit broadband electromagnetic radiation that ranges from radio to extremely high $\gamma$-ray energies. The jets of blazars are highly Doppler boosted, resulting in flux variations over the entire accessible electromagnetic wavebands. Flat spectrum radio quasars (FSRQs) and BL Lacertae (BL Lacs) objects are the two subclass of blazars. The difference between these subclasses is determined by the equivalent width (EW) of the emission lines in their optical spectra, with FSRQs having EW > 5 \AA{} and BL Lacs having EW < 5 \AA{} \citep{1991ApJS...76..813S}. A more physical criterion for distinguishing between FSRQs and BL Lacs was proposed by \cite{2011MNRAS.414.2674G} which is based on the ratio of broad line region (BLR) luminosity ($L_{BLR}$) to Eddington Luminosity ($L_{Edd}$). FSRQs have the value $L_{BLR}$/$L_{Edd}$ $>$ 5 $\times$ $10^{-4}$, whereas BL Lacs have the value \mbox{$L_{BLR}$/$L_{Edd}$ $<$ 5 $\times$ $10^{-4}$}.

The broad-band spectral energy distributions (SEDs) of blazars comprise of two humps; the low energy hump and the high energy hump. The low energy hump peaks at optical/UV/X-ray region and the high energy hump peaks at MeV/GeV/TeV region \citep{1998MNRAS.299..433F, 2010ApJ...716...30A, 2016ApJS..224...26M}. The genesis of the low energy hump is well understood through synchrotron emission mechanism of the relativistic electrons, while the high energy hump originates through inverse Compton (IC) emission process \citep{2010ApJ...716...30A}. Based on the position of the synchrotron peak frequency ($\nu_{syn}$), blazars are further classified as low synchrotron peaked (LSP; $\nu_{syn}$ $<$ $10^{14}$ Hz), intermediate synchrotron peaked (ISP; $10^{14}$ Hz $<$ $\nu_{syn}$ $<$ $10^{15}$ Hz) and high synchrotron peaked (HSP; $\nu_{syn}$ $>$ $10^{15}$ Hz) blazars.

Due to the Doppler boosting in blazars' jets, the observed emission $S_{obs}$ relative to the emission in the co-moving frame $S_{int}$ is defined as e.g., \citep[]{2017RAA....17...66L}
\begin{equation}
    S_{obs} = S_{int} \delta^{q} 
\end{equation}
 Where q = 3+$\alpha$ for a moving compact source and q = 2 + $\alpha$ for a stationary jet \citep{1985ApJ...295..358L}. {$\alpha$ is the spectral index which is defined 
 as $f_{\nu}$ $\propto$ $\nu^{-\alpha}$}. $\delta$ is the Doppler boosting factor which is described as $\delta$ = 1/$\Gamma$(1-$\beta$cos$\theta$), where $\Gamma$ is bulk Lorentz factor ($\Gamma$ = 1/$(1-\beta^{2})^{1/2}$), $\beta$ is the speed of jet in units of the speed of light and $\theta$ is the viewing angle between observer's line of sight and jet's axis. The observed time is also shortened by the effect of Doppler boosting by a factor $\delta^{-1}$. These two consequences of the Doppler boosting increase the chances of observed variations in blazars over a wide range of wavelengths and make them the brightest objects in the extragalactic sky.
 
 The study of flux variability is a valuable tool for determining the size of the emission zone in blazars. Blazars exhibit variability over a wide range of timescales, from few minutes to several years, across the full wavelength range i.e., from radio to $\gamma$-rays \citep{2012MNRAS.425.3002G, 2013A&A...552A..11R, 2017ApJ...841..123P, 2017ApJ...844...62P, 2018ApJ...854L..26S}. The variability in blazars can be explained by the shock-in-jet model \citep{1985ApJ...298..114M}. In this scenario, the inhomogeneities in the jet flow produce relativistic shocks. These shocks travel through the jet plasma at relativistic speed and accelerate the particles. The other scenario is the magnetic reconnection, which is responsible for the rapid variations in blazar jets and has been investigated extensively in recent years \citep{2013MNRAS.431..355G, 2015MNRAS.450..183S, 2016ApJ...816L...8W, 2020NatCo..11.4176S}. Within the reconnection region, magnetic reconnection possibly enables small-scale ultra-relativistic flow, also called jet-in-jet scenario \cite{2010MNRAS.402.1649G, 2013MNRAS.431..355G, 2020NatCo..11.4176S}. The ultra-relativistic motion of small plasmoids causes additional Doppler boosting and results in very short and bright flares. 
 
 The launch of $\it{Fermi}$ Telescope in 2008 gave us an unprecedented opportunity to explore the $\gamma$-ray regime in blazars. Using $\it{Fermi}$ data at $\gamma$-ray energies, flux doubling time scales have been detected as short as few minutes in both FSRQs and BL Lacs \citep{2007ApJ...664L..71A, 2013ApJ...762...92A, 2016ApJ...824L..20A, 2018ApJ...854L..26S}. Causality considerations in these circumstances point to highly compact $\gamma$-ray emission zones in blazars' jets. The detection of $\gamma$-ray photons with energy greater than 10 GeV in FSRQs leads to the conclusion that the $\gamma$-ray emission region should be located at outside the cavity formed by the broad emission line (BLR)~\citep{2006ApJ...653.1089L}. Otherwise, pair production on UV photons emitted by (BLR) clouds should severely attenuate the $\gamma$-ray photons. Despite several studies, the physics of $\gamma$-ray variability in blazars is still a captivating subject of research. 
 
 Ton 599 (4FGL J1159.5 + 2914; \citep{2020ApJS..247...33A}) is an FSRQ, located at redshift z = 0.725 \citep{2010MNRAS.405.2302H}  with R.A. = $11^{h} 59^{m} 31.8^{s}$ and Dec. = $+29^{\circ} 14{^\prime} 43.8{"}$. Ton 599 is a strongly polarized and a highly optically violent variable quasar~\citep{2006PASJ...58..797F}. It was first detected in high energy $\gamma$-rays by the Energetic Gamma Ray Experiment Telescope (EGRET) in the second EGRET catalogue~\citep{1995ApJS..101..259T} and later by the $\it{Fermi}$ Large Area Telescope ($\it{Fermi}$-LAT) \citep{2010ApJ...715..429A}. The correlation study between radio and $\gamma$-ray bands was carried out by \citep{2014MNRAS.445.1636R} for this source to put the constraints on $\gamma$-ray emission region in parsec-scale jets. In 2017, for the first time, Ton 599 went through a protracted flaring condition spanning the full electromagnetic spectrum. During this flare, the maximum $\gamma$-ray flux observed was 1.26 $\times$ $10^{-6}$ ph cm$^{-2}$ s$^{-1}$ \citep{2019ApJ...871..101P}. The detailed study of $\gamma$-ray flux variability during this flare was carried out by \citep{2018ApJ...866..102P, 2019ApJ...871..101P}.  A multiwavelength study and the broad-band SED modelling of this flare was conducted by \citep{2020MNRAS.492...72P} using a two-component leptonic emission model. An EC mechanism with a dusty torus (DT) photon field producing seed photons was identified to be responsible for the GeV emission in this study. Ton 599 recently displayed a bright flare from July to September 2021, allowing us to investigate its $\gamma$-ray emission process and its consequences during this source's flaring condition. 
 
 In this paper, we describe the $\gamma$-ray analysis performed on the source Ton 599 utilising data collected over a one-year period from September 2020 to August 2021, with the goal of constraining the $\gamma$-ray emission region in blazars. The data used in this study is presented in Section \ref{sec:2}. Section \ref{sec:3} describes the $\gamma$-ray light curve, followed by results in Section 4 and discussion in Section \ref{sec:5}. Section 6 presents a summary of the results.

\section{\boldmath{$\gamma$}-ray Observations and Data Reduction \label{sec:2}}
In this work, we used the $\gamma$-ray observations of Ton 599 taken by the $\it{Fermi}-$LAT for a period of $\sim$ 1 year, from 2020 September to 2021 August (MJD: 59,093 
--59,457). $\it{Fermi}-$LAT is a pair-conversion $\gamma$-ray telescope that can detect $\gamma$ rays with energies ranging from 20 MeV to more than 1 TeV. It has a large field of view of about 2.4 sr and  scans the full sky in every 3 h, covering 20\% of the sky at any time \citep{2009ApJ...697.1071A}. We used the package ScienceTools v1.2.23 with the instrument response function P8R3$\_$SOURCE$\_$V3\endnote{\url{https://fermi.gsfc.nasa.gov/ssc/data/analysis/documentation/}, accessed on 2021/09/01 
} for our analysis. We used the latest LAT Pass 8 data in the energy range 100 MeV to 300 GeV, where the photon-like events are classified as `evclass = 128, evtype = 3'. The region of interest (ROI) is specified as a circle with a radius of $10^{\circ}$ and is centred on the $\gamma$-ray position of the source. We used a maximum zenith angle of $90^{\circ}$ to remove $\gamma$-ray contamination from the earth's limb. The latest isotropic model 'iso$\_$P8R3$\_$SOURCE$\_$V3$\_$v1' and the Galactic diffuse emission model 'gll$\_$iem$\_$v07' were used to analyze the data. The recommended condition ‘(DATA QUAL $>$ 0)\&\&(LAT CONFIG= = 1)' was then used to construct the required good time intervals. Over the time period of interest, an unbinned likelihood analysis is performed to generate the 1-day binned $\gamma$-ray light curve of Ton 599. In the light curve, the source was considered to be detected if the test statistics (TS) $>$ 9, which corresponds to a 3$\sigma$ detection \citep{1996ApJ...461..396M}. Our final one-day binned $\gamma$-ray light curve contains 256 confirmed measurements of Ton 599.

\section{\boldmath{$\gamma$}-ray Light Curve \label{sec:3}}
\begin{figure}[H]
\widefigure
\includegraphics[width=18 cm]{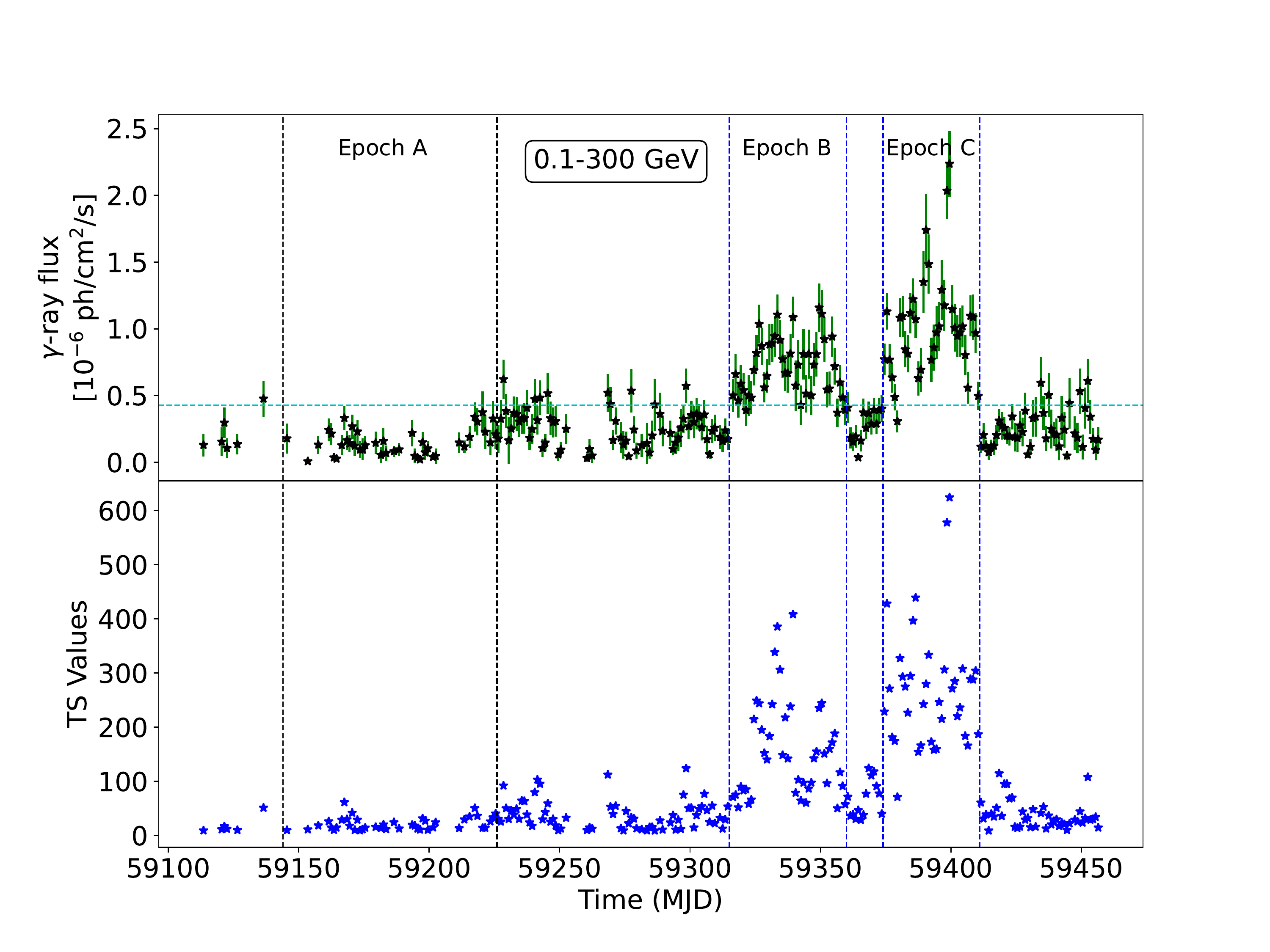}
\caption{One-day 
 binned $\gamma$-ray light curve of Ton 599 {(\textbf{upper panel}) with the TS values corresponding to each flux value (\textbf{lower panel}) }. The vertical black lines refer to the quiescent epoch (Epoch A), whereas vertical blue lines refer to the $\gamma$-ray flaring epochs (Epoch B and Epoch C). In the upper panel the horizontal cyan line represents the average flux from September 2020 to August 2021.  \label{fig1}}
\end{figure}

The one-day binned $\gamma$-ray light curve of Ton 599 from 2020 September to 2021 August (MJD: 59,093--59,457) is shown in Figure \ref{fig1}. For the entire period of the light curve, we estimated the average $\gamma$-ray flux which was found to be 0.43 $\pm$ 0.36 $\times$ $10^{-6}$ ph cm$^{-2}$ s$^{-1}$. Based on the average $\gamma$-ray flux, we visually identified three different flux states, namely, quiescent state, pre-flaring state, and the main flaring state, which were labelled as epoch A, epoch B, and epoch C, respectively. The duration of these  epochs are marked by vertical lines in Figure \ref{fig1} and the details of these epochs are given in Table \ref{tab1}. We classified epoch A as a quiescent epoch since the flux was lower than the average flux for the whole time span. During epochs B and C, the flux increased 2--3 times than the average flux, so we classified these epochs as flaring epochs. Over the whole time span of the light curve, the value of the flux was maximum during the flaring epoch C. The value of the maximum flux was determined to be 2.24 $\pm$ 0.25 $\times$ $10^{-6}$ ph cm$^{-2}$ s$^{-1}$ at MJD 59,399.50.

\begin{figure}[H]
\widefigure
\vbox{
\hbox{
\includegraphics[scale=0.45]{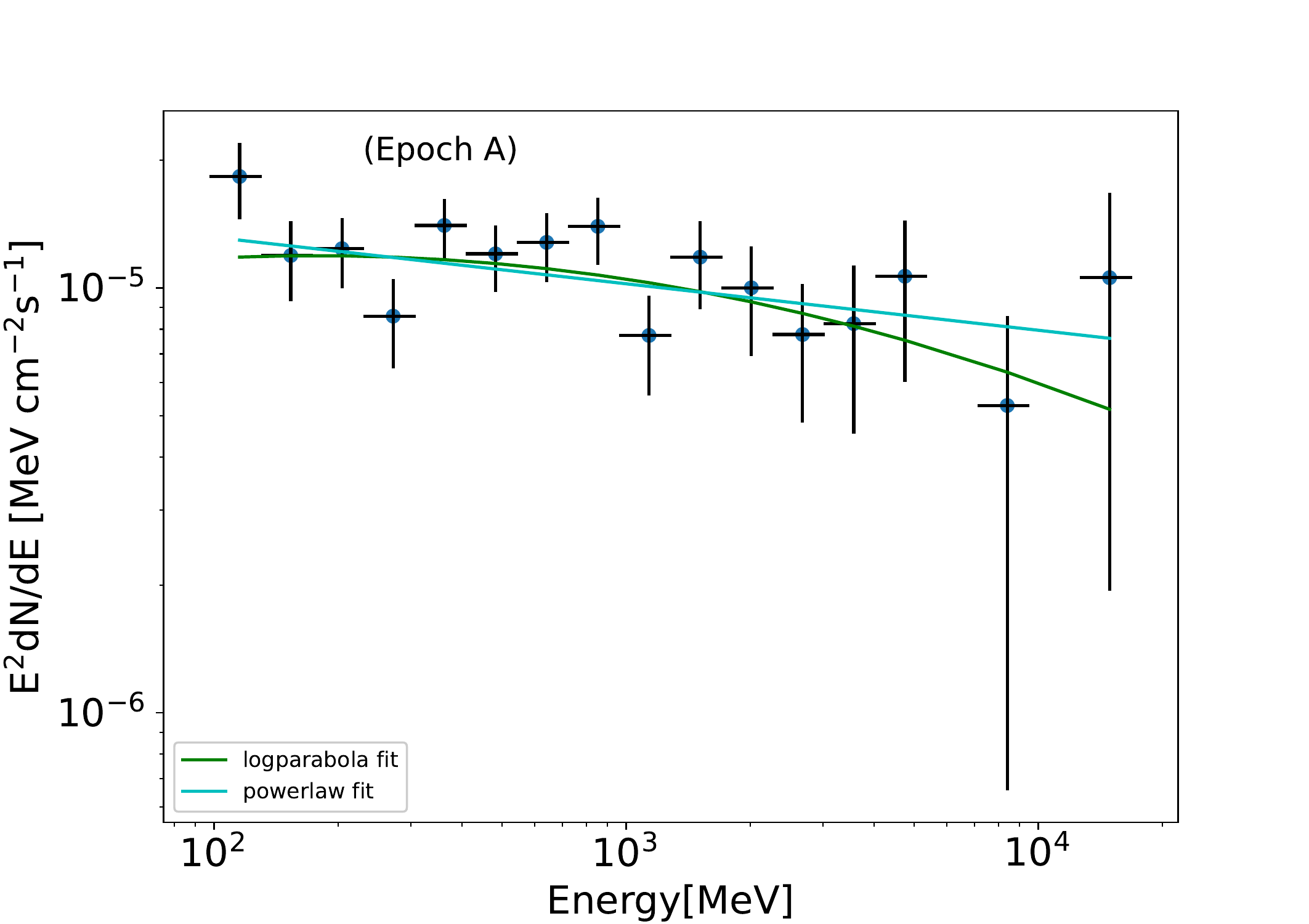}
\includegraphics[scale=0.45]{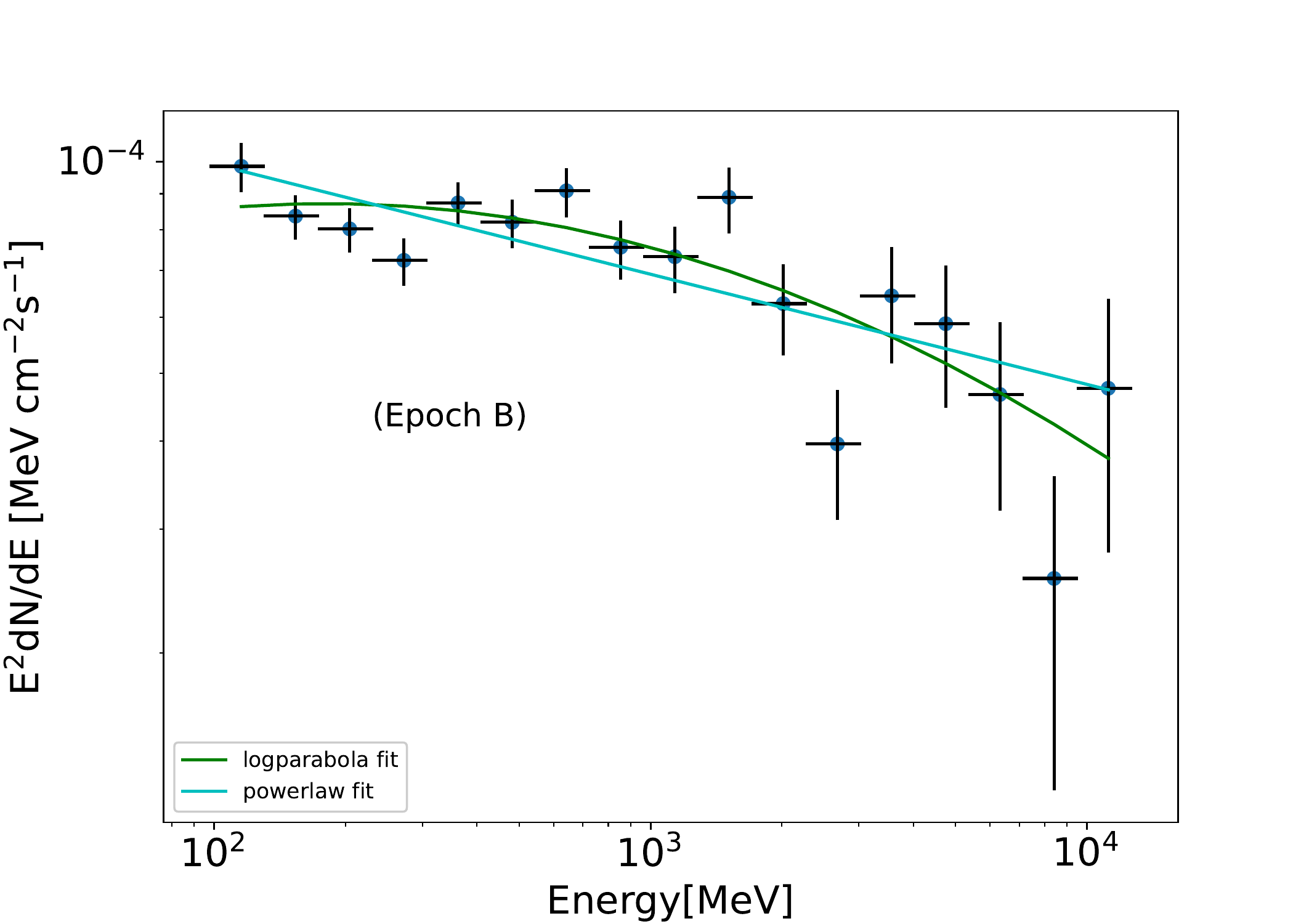}}
\hbox{\hspace{5cm}
\includegraphics[scale=0.45]{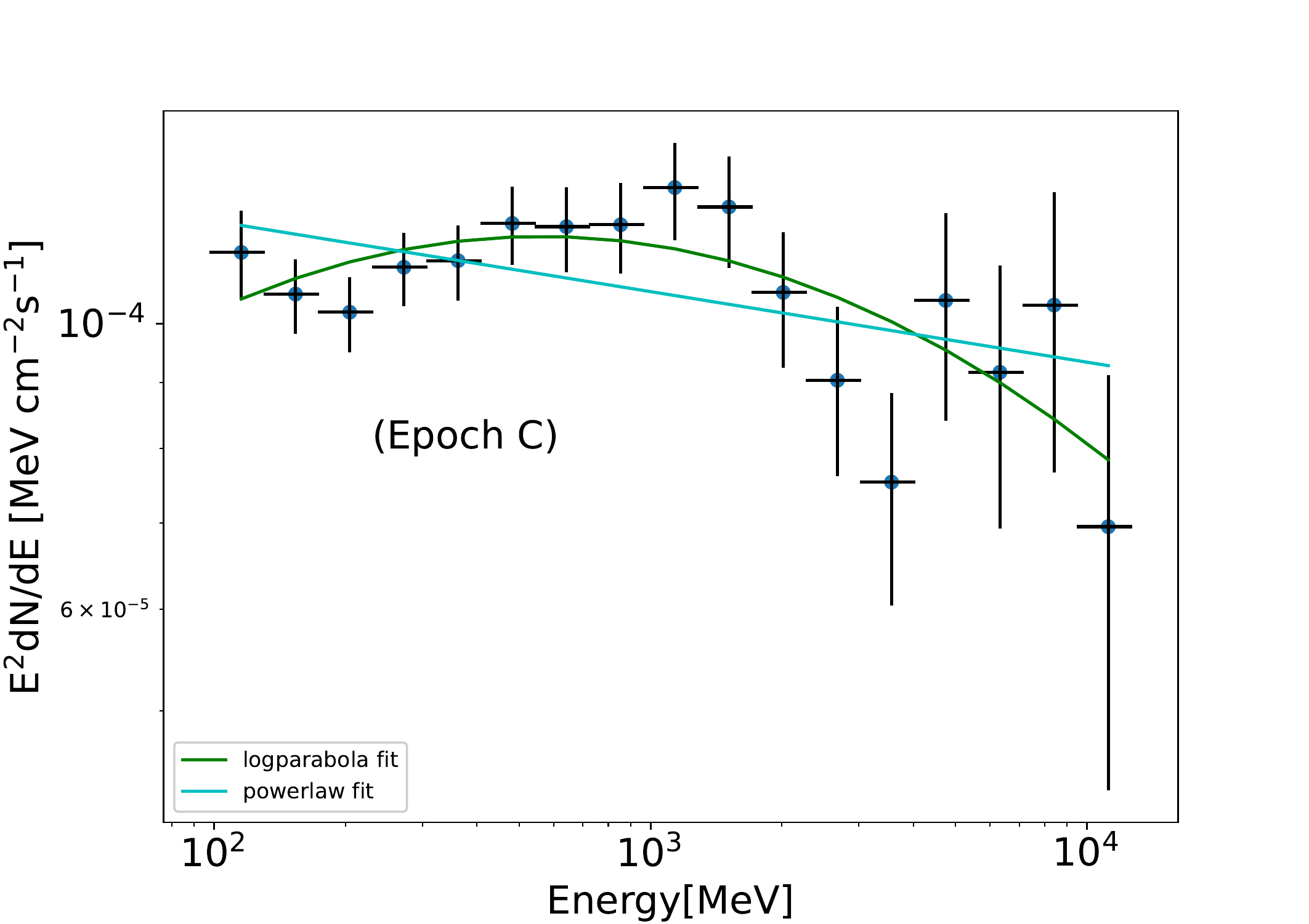}}}
\caption{\label{figure-2}$\gamma$-ray spectrum for the epochs (A--C). The name of the epoch is mentioned in each plot.}
\end{figure}

\end{paracol}
\nointerlineskip
    \begin{specialtable}[H]
    \tablesize{\footnotesize}
    \widetable
\caption{The epochs considered for the $\gamma$-ray light curve study are listed in the table below. {Here, N, $\bar{x}$, and $F_{var}$ denote the total number of data points, the average $\gamma$-ray flux, and  the fractional variability amplitude, respectively, for the epoch.} The $\gamma$-ray flux is in the units of $10^{-6}$ ph cm$^{-2}$ s$^{-1}$. 
\label{tab1}}
\setlength{\cellWidtha}{\columnwidth/10-2\tabcolsep+0.0in}
\setlength{\cellWidthb}{\columnwidth/10-2\tabcolsep+0.0in}
\setlength{\cellWidthc}{\columnwidth/10-2\tabcolsep-0.0in}
\setlength{\cellWidthd}{\columnwidth/10-2\tabcolsep-0.0in}
\setlength{\cellWidthe}{\columnwidth/10-2\tabcolsep-0.0in}
\setlength{\cellWidthf}{\columnwidth/10-2\tabcolsep-0.0in}
\setlength{\cellWidthg}{\columnwidth/10-2\tabcolsep-0.3in}
\setlength{\cellWidthh}{\columnwidth/10-2\tabcolsep-0in}
\setlength{\cellWidthi}{\columnwidth/10-2\tabcolsep-0in}
\setlength{\cellWidthj}{\columnwidth/10-2\tabcolsep+.3in}
\scalebox{1}[1]{\begin{tabularx}{\columnwidth}{>{\PreserveBackslash\centering}p{\cellWidtha}>{\PreserveBackslash\centering}p{\cellWidthb}>{\PreserveBackslash\centering}p{\cellWidthc}>{\PreserveBackslash\centering}p{\cellWidthd}>{\PreserveBackslash\centering}m{\cellWidthe}>{\PreserveBackslash\centering}m{\cellWidthf}>{\PreserveBackslash\centering}m{\cellWidthg}>{\PreserveBackslash\centering}m{\cellWidthh}>{\PreserveBackslash\centering}m{\cellWidthi}>{\PreserveBackslash\centering}m{\cellWidthj}}
\toprule
 \multirow{2}{*}{\textbf{Epochs}\vspace{-4pt}}    &\multicolumn{2}{c}{\textbf{MJD}} & \multicolumn{2}{c}{\textbf{Calendar date (dd-mm-yyyy)}}& \multirow{2}{*}{\textbf{Peak Flux}\vspace{-4pt}}&\multirow{2}{*}{\boldmath{$N$}\vspace{-4pt}} &\multirow{2}{*}{\boldmath{$\bar{x}$}\vspace{-4pt}} &\multirow{2}{*}{\boldmath{$F_{var}$}\vspace{-4pt}} & \multirow{2}{*}{\textbf{Remarks}\vspace{-4pt}}\\ 
     \cmidrule{2-5}
           & \textbf{Start} & \textbf{End}  & \textbf{Start} & \textbf{End}  &  &  &  &  &   
\\
\midrule
Epoch A		& 59,144 & 59,226 & 22-10-2020 & 12-01-2021 & 0.37 $\pm$ 0.16 & 42 & 0.15 $\pm$ 0.01 & 0.36 $\pm$ 0.12	& Quiescent \\
Epoch B		& 59,315 & 59,360 & 11-04-2021 & 24-05-2021 & 1.16 $\pm$ 0.18 & 44 & 0.71 $\pm$ 0.02 & 0.22 $\pm$ 0.04	& Flaring\\
Epoch C     & 59,374 & 59,411 & 09-06-2021 & 16-07-2021 & 2.24 $\pm$ 0.25 & 37 & 1.02 $\pm$ 0.03 & 0.35 $\pm$ 0.03   & Flaring\\
\hline
\end{tabularx}}
\end{specialtable}
\begin{paracol}{2}
\switchcolumn

\clearpage
\end{paracol}
\nointerlineskip
\vspace{-6pt}
\begin{paracol}{2}
\switchcolumn

\section{Results}
\subsection{$\gamma$-ray Flux Variability}
We used the fractional variability amplitude (F$_{var}$) to characterize the flux variability properties of Ton 599. The F$_{var}$ is commonly used to quantify the intrinsic variations in blazar light curves and is defined as, e.g., \citep[]{2003MNRAS.345.1271V, 2017ApJ...841..123P,2018ApJ...859...49P}
\begin{equation}
F_{var} = \sqrt{\frac{S^{2} - \overline{\sigma^{2}_{err}}}{\bar{x}^{2}}}
\end{equation}

In the above equation $S^{2}$ is the sample variance and $\overline{\sigma^{2}_{err}}$ is mean square error. These are defined as
\begin{equation}
S^{2} = \frac{1}{N-1}\sum_{i=1}^{N}(x_{i} - \bar{x})^{2}
\end{equation}
and
\begin{equation}
\overline{\sigma^{2}_{err}} = \frac{1}{N}\sum_{i=1}^{N}{\sigma^{2}_{err,i}} \,
\end{equation}

Here N is the total number of data points in the light curve and $\bar{x}$ is the arithmetic mean of the light curve.

The uncertainty in $F_{var}$ is given by the following equation
\begin{equation}
err(F_{var}) = \sqrt{\Bigg(\sqrt{\frac{1}{2N}}\frac{\overline{\sigma_{err}^{2}}}{\bar{x}^{2}F_{var}}\Bigg)^{2} + \Bigg(\sqrt{\frac{\overline{\sigma_{err}^{2}}}{N}}\frac{1}{\bar{x}}\Bigg)^{2}}
\end{equation}

{For each epoch, we calculated the value of $F_{var}$ separately by considering the average flux ($\bar{x}$) and the data points ($N$) within the epoch only. The values of $N$, $\bar{x}$, and F$_{var}$ for each epoch are listed in} Table \ref{tab1}. The errors calculated using equation (5) are given in Table \ref{tab1}. For epoch A, the error in the value of $F_{var}$ is relatively large which also indicates that it is a quiescent state. We considered the source to be variable in epochs if $F_{var}$ $>$ 3 $\times$ err($F_{var}$). According to our criteria for variations, the source was not variable in epoch A (quiescent state), while significant $\gamma$-ray flux variations were noticed  during epochs B and C with F$_{var}$ values of 22\% and 35\%, respectively. 

\subsection{$\gamma$-ray Spectral Fitting}
We modeled the $\gamma$-ray spectrum of Ton 599 for each epoch to investigate the inherent distribution of electrons that causes the $\gamma$-ray emission during the epoch. We used the power-law (PL) and the log parabola (LP) models to fit the $\gamma$-ray spectra of Ton 599 {using the maximum likelihood analysis}. The PL model is defined as follows \citep{2012ApJS..199...31N}:
\begin{equation}
\frac{dN(E)}{dE}=N_{\circ}\left(\frac{E}{E_{\circ}}\right)^{\Gamma}
\end{equation}
where $N_{\circ}$ is the prefactor (normalization of the energy spectrum), $E_{\circ}$ is the {pivot energy ($\sim$523.51 MeV) given in the 4FGL catalog,} and $\Gamma$ is the photon index, while, the LP model is given by
\begin{equation}
\frac{dN(E)}{dE}=N_{\circ}\left(\frac{E}{E_{b}}\right)^{-\alpha-\beta ln\left(\frac{E}{E_{b}}\right)}
\end{equation}
In this equation $N_{\circ}$ is the normalization, $E_{b}$ is {the pivot energy (same as in PL)}, $\alpha$ is photon index at $E_{b}$ and $\beta$ is the curvature index that defines the curvature of the spectrum around the peak.

For each epoch, the model fitted $\gamma$-ray spectra of Ton 599 is shown in Figure \ref{figure-2}, and the values of best fitted model parameters are given in Table \ref{table-2}. {In Figure \ref{figure-2}, the uncertainties are large at higher energies because of relatively low photon counts at these energies.}

To determine whether the $\gamma$-ray spectrum has a curvature or not, and which model (LP or PL) best describes the $\gamma$-ray spectrum of Ton 599, we calculated TS$_{curve}$ (curvature of the test statistics) \cite{2012ApJS..199...31N}, which is defined as:
\begin{equation}
TS_{curve} = 2(log L_{LP} - log L_{PL})
\end{equation}
In the above equation, L represents the likelihood function. The value of TS$_{curve}$ for each epoch is given in Table \ref{table-2}. We employed the $TS_{curve}$ $>$ 16 threshold (i.e., 4$\sigma$ level; \citep{1996ApJ...461..396M}) to determine the presence of statistically significant curvature in the $\gamma$-ray spectrum. We found that the LP model best describes the $\gamma$-ray spectra of Ton 599 for all three epochs.
\end{paracol}
\nointerlineskip
\vspace{-6pt}
%
    \begin{specialtable}[H]
    \tablesize{\footnotesize}
    \widetable
\caption{\label{table-2}Details of the PL and LP model fits for the three epochs. Here, $\gamma$-ray flux is in the units of $10^{-6}$ ph cm$^{-2}$ s$^{-1}$. 
}
\setlength{\cellWidtha}{\columnwidth/9-2\tabcolsep-.2in}
\setlength{\cellWidthb}{\columnwidth/9-2\tabcolsep+0.2in}
\setlength{\cellWidthc}{\columnwidth/9-2\tabcolsep-0.0in}
\setlength{\cellWidthd}{\columnwidth/9-2\tabcolsep-0.0in}
\setlength{\cellWidthe}{\columnwidth/9-2\tabcolsep-0.0in}
\setlength{\cellWidthf}{\columnwidth/9-2\tabcolsep-0.0in}
\setlength{\cellWidthg}{\columnwidth/9-2\tabcolsep-0in}
\setlength{\cellWidthh}{\columnwidth/9-2\tabcolsep-0in}
\setlength{\cellWidthi}{\columnwidth/9-2\tabcolsep-0in}
\scalebox{1}[1]{\begin{tabularx}{\columnwidth}{>{\PreserveBackslash\centering}p{\cellWidtha}>{\PreserveBackslash\centering}p{\cellWidthb}>{\PreserveBackslash\centering}p{\cellWidthc}>{\PreserveBackslash\centering}p{\cellWidthd}>{\PreserveBackslash\centering}m{\cellWidthe}>{\PreserveBackslash\centering}m{\cellWidthf}>{\PreserveBackslash\centering}m{\cellWidthg}>{\PreserveBackslash\centering}m{\cellWidthh}>{\PreserveBackslash\centering}m{\cellWidthi}}
\toprule

     \multirow{2}{*}{\textbf{Epoch}\vspace{-4pt}} & \multicolumn{3}{c}{\textbf{PL}} & \multicolumn{4}{c}{\textbf{LP}} & \multirow{2}{*}{\textbf{TS}\boldmath{$_{curve}$}\vspace{-4pt}}\\ 
    \cmidrule{2-8}
           & \boldmath{$\Gamma$} & \textbf{Flux}  & \boldmath{$-$}\textbf{Log L}  & \boldmath{$\alpha$} & \boldmath{$\beta$} & \textbf{Flux} & \boldmath{$-$}\textbf{Log L} &\\
\midrule
  A          &$-$2.381 $\pm$ 0.004&0.135 $\pm$ 0.002 & 1,46,206.160 & 2.045 $\pm$ 0.085 & 1.075$\pm$ 0.473 &  0.111 $\pm$ 0.012 & 1,46,191.326 & 29.67 \\
  B          &$-$2.142 $\pm$ 0.001&0.800 $\pm$ 0.002 & 92,649.820 & 2.051 $\pm$ 0.021 & 0.973$\pm$ 0.117 & 0.735 $\pm$ 0.020  & 92,634.266  & 31.108 \\
  C          &$-$2.044 $\pm$ 0.002&1.084 $\pm$ 0.005 & 84,267.700 & 1.923 $\pm$ 0.002 & 0.958$\pm$ 0.011 & 0.980 $\pm$ 0.003& 84,245.258 &  44.884 \\
\bottomrule
\end{tabularx}}
    \end{specialtable}
\vspace{-10pt}
\begin{paracol}{2}
\switchcolumn

\subsection{Location of the $\gamma$-ray Emission Region}
To determine the size of the $\gamma$-ray emitting region, the flux doubling time scale is usually estimated, e.g.,~\citep[]{2015ApJ...808L..48P, 2019ApJ...871..101P}. We calculated the flux doubling time scale for the one-day binned $\gamma$-ray light curve during the main-flaring epoch (epoch C) of Ton 599 as follows:
\begin{equation}
    F(t_{2}) = F(t_{1})\times2^{\Delta t/\tau_{d}}
\end{equation}
In the above expression, \emph{F}($t_{1}$) and \emph{F}($t_{2}$) are the flux values at times $t_{1}$ and $t_{2}$ respectively, $\Delta$~\emph{t} = $t_{2}-t_{1}$ and $\tau_{d}$ represents the flux doubling time. 

We found a flux doubling timescale of $\sim$13.2 h during the epoch C for the blazar Ton 599. Using the flux doubling time scale, we estimated the size of the $\gamma$-ray emitting region for Ton 599 using the following expression:
\begin{equation}
    r \leq c \tau_{d} \delta / (1+z)
\end{equation}

{A gamma-ray Doppler factor of delta = 12.5 was calculated for Ton 599 by \citep{2002PASJ...54..159Z} using the multiwavelength data and the known radio Doppler factors with the assumption that the observed boosted emission is from the SSC model. This value is consistent with the lower limits (10.75 and 13.45) of delta obtained for the two bright flares of Ton 599 by \cite{2018ApJ...866..102P}}. Using $\delta$ = 12.5, the size of the $\gamma$-ray emission region is estimated to be 1.03 $\times$ $10^{16}$ cm. 

According to \citep{2011A&A...530A..77F}, the location of the $\gamma$-ray emission region with respect to the central super massive black hole (SMBH) can be approximated by using the relation R = r/$\phi$, where r is the size of the $\gamma$-ray emitting blob and {$\phi$ is the jet opening angle.} 

{An intrinsic opening angle of $0.58^{\circ}$ ($\sim$ 0.01 radian) was determined for Ton 599 by \cite{2009A&A...507L..33P}. They estimated the jet opening angles for Fermi detected AGNs using 15.4 GHz Very Long Baseline Array (VLBA) observations following two different methods: (a) by analyzing transverse jet profiles in the image plane and (b) by model fitting the data in the (u, v) plane.} {Using these values of $r$ and $\phi$, the location of the $\gamma$-ray emitting region for Ton 599 was found to be at 1.03 $\times$ $10^{18}$ cm from the SMBH.}

\section{Discussion \label{sec:5}}
In this section we give our interpretation of the findings of above-mentioned analyses, as well as a discussion of them.
\subsection{$\gamma$-ray Flux Variability}
 In the $\gamma$-ray band, blazars exhibit remarkable flux variability. For all the three epochs of the source Ton 599, we estimated the flux variability amplitude using one-day binned light curves. We found that during the main-flaring epoch (epoch C), the source showed largest variations with $F_{var}$ = 0.35 $\pm$ 0.03. During the pre-flaring epoch (epoch B), the source was also variable with $F_{var}$ = 0.22 $\pm$ 0.04 and during the quiescent epoch (epoch A), the source was not significantly variable ($F_{var}$ = 0.36 $\pm$ 0.12) within the error bar. During the main-flare (epoch C), the source showed a maximum flux of \mbox{2.24 $\pm$ 0.25 $\times$ $10^{-6}$ ph cm$^{-2}$ s$^{-1}$} at MJD 59,399.50, which is larger than the maximum flux observed during  the 2017 flare~\cite{2019ApJ...871..101P}. We found that the photon index value is 1.99 $\pm$ 0.10 during the main-flaring epoch (epoch C), when flux is highest, harder than the 4FGL value of 2.19 $\pm$ 0.01.
 
 The observed $\gamma$-ray variability in the blazars could be attributed to both intrinsic and extrinsic factors. The distribution of relativistic electrons responsible for the emission is one of the intrinsic effects. These relativistic electrons, which can be accelerated to a Lorentz factor of up to $10^{6}$, are responsible for non-thermal emission from blazar jets via synchrotron or inverse Compton emission processes. The distribution of these relativistic electrons and seed photon responsible for the inverse Compton emission process (synchrotron self Compton (SSC); \citep{1985ApJ...298..114M, 1989ApJ...340..181G} and external Compton (EC); \citep{1987ApJ...322..650B,1994ApJ...421..153S}) are the intrinsic factors that cause short-term variability in jets. Extrinsic factors, such as the moving blob's high bulk Lorentz factor ($\Gamma$) $\sim$ 50, are in addition to intrinsic factors. Though the large bulk Lorentz factor is a favourable explanation in the case of BL Lacs, because the seed photons come from inside the jets and produce $\gamma$-rays through inverse Compton. On the other hand, the seed photons in FSRQ, emanate from outside the jets and produce $\gamma$-rays. However, pair-production through dense broad line region (BLR) can attenuate $\gamma$-ray emission. As a result, in the case of FSRQs, the large bulk Lorentz factor that causes $\gamma$-ray variability is not a plausible explanation \citep{2011AdSpR..48..998S}. The $\gamma$-ray flux variability in blazars jets can be explained by the magnetic reconnection (jet-in-jet) scenario, in which the mini jets generated in the jets can have a large bulk Lorentz factor and they can produce $\gamma$-ray flux variability on short time scale \cite{2013MNRAS.431..355G, 2020NatCo..11.4176S}.
   
\subsection{$\gamma$-ray Spectra}
For each epoch, we modelled the $\gamma$-ray spectra of Ton 599 using the power law and the log parabola models. We used the TS$_{curve}$ value to determine the best fitted model. Based on the TS$_{curve}$ value, we found that the LP model best describes the $\gamma$-ray spectra of Ton 599 during all the three epochs indicating that the GeV $\gamma$-ray spectrum of Ton 599 is curved. 

FSRQs usually have a curved GeV $\gamma$-ray spectrum, e.g.,~\cite[]{2010ApJ...710.1271A,2015ApJ...803...15P,2019MNRAS.486.1781R}. Several theories have been proposed, in the literature, to explain the curved $\gamma$-ray spectra of FSRQs. The attenuation of $\gamma$-ray photons by pair-production (inside the BLR) could explain the curved $\gamma$-ray spectrum \citep{2016MNRAS.458..354C}. A viable scenario for defining the curve in the $\gamma$-ray spectrum is the Klein-Nishina effect on the inverse Compton scattering of BLR photons through relativistic electrons present in the jet \citep{2013ApJ...771L...4C}. However, several investigations conducted to locate the $\gamma$-ray emission region discovered that the observed $\gamma$-ray spectra in FSRQs are not caused by IC scattering of BLR photons and that the $\gamma$-ray emission site is located outside the BLR~\citep{2018MNRAS.477.4749C, 2020NatCo..11.4176S}. The curved $\gamma$-ray spectra of FSRQs could also be explained by the curved energy distribution of the electrons emitting the radiation \citep{2009A&A...501..879T, 2015ApJ...809..174D}.

   
\subsection{Location of the $\gamma$-ray Emitting Region}
We calculated the minimum $\gamma$-ray flux doubling timescale of 13.2 h during epoch C for Ton 599. Using this minimum doubling timescale, we estimated the size of the $\gamma$-ray emitting region as 1.03 $\times$ $10^{16}$ cm. We have also identified the location of the $\gamma$-ray emitting blob to be at a distance of {1.03 $\times$ $10^{18}$} cm from the SMBH. {The size of the BLR for TON 599 was estimated as $\sim$ (2.11$-$2.45) $\times$ $10^{17}$ cm by \cite{2018ApJ...866..102P}.} So, the location of the $\gamma$-ray emitting blob, found in this work, is outside of the BLR. {During the 2017 $\gamma$-ray flare of this source, using two component leptonic model, Ref. \cite{2020MNRAS.492...72P} found that the seed photons for the GeV emission produced by the dusty torus (DT) and the GeV emitting region is located outside of the BLR. Our results are consistent with the findings of \cite{2020MNRAS.492...72P}.} 

\section{Summary}
We investigated the $\gamma$-ray flux and spectral variability of the blazar Ton 599 during MJD 59,093 to MJD 59,457. 
 For our study, we have chosen three epochs of different flux states: quiescent, pre-flaring, and main-flaring. The outcomes of the analysis of these epochs are summarised below.
 \begin{itemize}
     \item We estimated the flux variability amplitude for the specified epochs. The largest $\gamma$-ray flux variations were found for the main-flaring epoch (C) with $F_{var}$ = 0.35 $\pm$ 0.03. The source also showed variations in epoch B with $F_{var}$ = 0.22 $\pm$ 0.04. However, no significant flux variation was observed in the quiescent epoch. 
     \item The $\gamma$-ray spectra were well fit by the LP model for all the three epochs. 
     \item {We estimated the size of the $\gamma$-ray emitting region as 1.03 $\times$ $10^{16}$ cm and the location of the $\gamma$-ray emitting blob could be outside of the BLR}.
 \end{itemize}
 
\vspace{6pt}
\authorcontributions{Conceptualization, Bhoomika Rajput; Data curation, Bhoomika Rajput and Ashwani Pandey; Formal analysis, Bhoomika Rajput; Investigation, Bhoomika Rajput; Project administration, Ashwani Pandey; Resources, Bhoomika Rajput; Visualization, Bhoomika Rajput; Writing -- original draft, Bhoomika Rajput; Writing -- review \& editing, Ashwani Pandey.}

\funding{This research received no external funding.}

\dataavailability{The data used in this work is publicly available from the Fermi-LAT data archive \url{https://fermi.gsfc.nasa.gov/ssc/data/access/}, accessed on 2021/09/01.} 

\acknowledgments{We thank the anonymous referees for their constructive comments, which helped us to improve the manuscript. This publication made use of data from the $\it{Fermi}$ Gamma-ray Space Telescope, which accessed from the Fermi Science Support Center. We also acknowledge the use of High Performance Computing Facility (Nova cluster) at Indian Institute of Astrophysics.}

\conflictsofinterest{The authors declare no conflict of interest.} 

\end{paracol}
\printendnotes[custom]

\reftitle{References}


\externalbibliography{no}


\begin{thebibliography}{999}

\bibitem[{Urry} and {Padovani}(1995)]{1995PASP..107..803U}
{Urry}, C.M.; {Padovani}, P.
\newblock {Unified Schemes for Radio-Loud Active Galactic Nuclei}.
\newblock {\em \pasp} {\bf 1995}, {\em 107},~803.
\newblock
 https://doi.org/10.1086/133630.

\bibitem[{Padovani} \em{et~al.}(2017){Padovani}, {Alexander}, {Assef}, {De
  Marco}, {Giommi}, {Hickox}, {Richards}, {Smol{\v{c}}i{\'c}},
  {Hatziminaoglou}, {Mainieri}, and {Salvato}]{2017A&ARv..25....2P}
{Padovani}, P.; {Alexander}, D.M.; {Assef}, R.J.; {De Marco}, B.; {Giommi}, P.;
  {Hickox}, R.C.; {Richards}, G.T.; {Smol{\v{c}}i{\'c}}, V.; {Hatziminaoglou},
  E.; {Mainieri}, V.; {Salvato}, M.
\newblock {Active galactic nuclei: what's in a name?}
\newblock {\em \aapr} {\bf 2017}, {\em 25},~2. {https://doi.org/10.1007/s00159-017-0102-9}.

\bibitem[{Stocke} \em{et~al.}(1991){Stocke}, {Morris}, {Gioia}, {Maccacaro},
  {Schild}, {Wolter}, {Fleming}, and {Henry}]{1991ApJS...76..813S}
{Stocke}, J.T.; {Morris}, S.L.; {Gioia}, I.M.; {Maccacaro}, T.; {Schild}, R.;
  {Wolter}, A.; {Fleming}, T.A.; {Henry}, J.P.
\newblock {The Einstein Observatory Extended Medium-Sensitivity Survey. II. The
  Optical Identifications}.
\newblock {\em \apjs} {\bf 1991}, {\em 76},~813.
\newblock
 {https://doi.org/10.1086/191582}.

\bibitem[{Ghisellini} \em{et~al.}(2011){Ghisellini}, {Tavecchio}, {Foschini},
  and {Ghirland a}]{2011MNRAS.414.2674G}
{Ghisellini}, G.; {Tavecchio}, F.; {Foschini}, L.; {Ghirland, A}.G.
\newblock {The transition between BL Lac objects and flat spectrum radio
  quasars}.
\newblock {\em \mnras} {\bf 2011}, {\em 414},~2674--2689.
  {https://doi.org/10.1111/j.1365-2966.2011.18578.x}.

\bibitem[{Fossati} \em{et~al.}(1998){Fossati}, {Maraschi}, {Celotti},
  {Comastri}, and {Ghisellini}]{1998MNRAS.299..433F}
{Fossati}, G.; {Maraschi}, L.; {Celotti}, A.; {Comastri}, A.; {Ghisellini}, G.
\newblock {A unifying view of the spectral energy distributions of blazars}.
\newblock {\em \mnras} {\bf 1998}, {\em 299},~433--448.
{https://doi.org/10.1046/j.1365-8711.1998.01828.x}.

\bibitem[{Abdo} \em{et~al.}(2010){Abdo}, {Ackermann}, {Agudo}, {Ajello},
  {Aller}, {Aller}, {Angelakis}, {Arkharov}, {Axelsson}, {Bach}, and
  et~al.]{2010ApJ...716...30A}
{Abdo}, A.A.; {Ackermann}, M.; {Agudo}, I.; {Ajello}, M.; {Aller}, H.D.;
  {Aller}, M.F.; {Angelakis}, E.; {Arkharov}, A.A.; {Axelsson}, M.; {Bach}, U.;
  et~al.
\newblock {The Spectral Energy Distribution of Fermi Bright Blazars}.
\newblock {\em \apj} {\bf 2010}, {\em 716},~30--70.
{https://doi.org/10.1088/0004-637X/716/1/30}.

\bibitem[{Mao} \em{et~al.}(2016){Mao}, {Urry}, {Massaro}, {Paggi},
  {Cauteruccio}, and {K{\"u}nzel}]{2016ApJS..224...26M}
{Mao}, P.; {Urry}, C.M.; {Massaro}, F.; {Paggi}, A.; {Cauteruccio}, J.;
  {K{\"u}nzel}, S.R.
\newblock {A Comprehensive Statistical Description of Radio-through-Gamma-Ray
  Spectral Energy Distributions of All Known Blazars}.
\newblock {\em \apjs} {\bf 2016}, {\em 224},~26.
{https://doi.org/10.3847/0067-0049/224/2/26}.

\bibitem[{Lin} \em{et~al.}(2017){Lin}, {Fan}, and {Xiao}]{2017RAA....17...66L}
{Lin}, C.; {Fan}, J.H.; {Xiao}, H.B.
\newblock {The intrinsic {\ensuremath{\gamma}}-ray emissions of Fermi blazars}.
\newblock {\em Res. Astron. Astrophys.} {\bf 2017}, {\em
  17},~066. {https://doi.org/10.1088/1674-4527/17/7/66}.

\bibitem[{Lind} and {Blandford}(1985)]{1985ApJ...295..358L}
{Lind}, K.R.; {Blandford}, R.D.
\newblock {Semidynamical models of radio jets: Relativistic beaming and source
  counts.}
\newblock {\em \apj} {\bf 1985}, {\em 295},~358--367. {https://doi.org/10.1086/163380}.

\bibitem[{Gaur} \em{et~al.}(2012){Gaur}, {Gupta}, {Strigachev}, {Bachev},
  {Semkov}, {Wiita}, {Peneva}, {Boeva}, {Slavcheva-Mihova}, {Mihov}, {Latev},
  and {Pandey}]{2012MNRAS.425.3002G}
{Gaur}, H.; {Gupta}, A.C.; {Strigachev}, A.; {Bachev}, R.; {Semkov}, E.;
  {Wiita}, P.J.; {Peneva}, S.; {Boeva}, S.; {Slavcheva-Mihova}, L.; {Mihov},
  B.; {Latev}, G.; {Pandey}, U.S.
\newblock {Optical flux and spectral variability of blazars}.
\newblock {\em \mnras} {\bf 2012}, {\em 425},~3002--3023.
{https://doi.org/10.1111/j.1365-2966.2012.21583.x}.

\bibitem[{Rani} \em{et~al.}(2013){Rani}, {Krichbaum}, {Fuhrmann},
  {B{\"o}ttcher}, {Lott}, {Aller}, {Aller}, {Angelakis}, {Bach}, {Bastieri},
  {Falcone}, {Fukazawa}, {Gabanyi}, {Gupta}, {Gurwell}, {Itoh}, {Kawabata},
  {Krips}, {L{\"a}hteenm{\"a}ki}, {Liu}, {Marchili}, {Max-Moerbeck},
  {Nestoras}, {Nieppola}, {Quintana-Lacaci}, {Readhead}, {Richards}, {Sasada},
  {Sievers}, {Sokolovsky}, {Stroh}, {Tammi}, {Tornikoski}, {Uemura},
  {Ungerechts}, {Urano}, and {Zensus}]{2013A&A...552A..11R}
{Rani}, B.; {Krichbaum}, T.P.; {Fuhrmann}, L.; {B{\"o}ttcher}, M.; {Lott}, B.;
  {Aller}, H.D.; {Aller}, M.F.; {Angelakis}, E.; {Bach}, U.; {Bastieri}, D.;
  et~al.
\newblock {Radio to gamma-ray variability study of blazar S5 0716+714}.
\newblock {\em \aap} {\bf 2013}, {\em 552},~A11.
{https://doi.org/10.1051/0004-6361/201321058}.

\bibitem[{Pandey} \em{et~al.}(2017){Pandey}, {Gupta}, and
  {Wiita}]{2017ApJ...841..123P}
{Pandey}, A.; {Gupta}, A.C.; {Wiita}, P.J.
\newblock {X-Ray Intraday Variability of Five TeV Blazars with NuSTAR}.
\newblock {\em \apj} {\bf 2017}, {\em 841},~123.
{https://doi.org/10.3847/1538-4357/aa705e}.

\bibitem[{Prince} \em{et~al.}(2017){Prince}, {Majumdar}, and
  {Gupta}]{2017ApJ...844...62P}
{Prince}, R.; {Majumdar}, P.; {Gupta}, N.
\newblock {Long-term Study of the Light Curve of PKS 1510-089 in GeV Energies}.
\newblock {\em \apj} {\bf 2017}, {\em 844},~62.
{https://doi.org/10.3847/1538-4357/aa78f4}.

\bibitem[{Shukla} \em{et~al.}(2018){Shukla}, {Mannheim}, {Patel}, {Roy},
  {Chitnis}, {Dorner}, {Rao}, {Anupama}, and {Wendel}]{2018ApJ...854L..26S}
{Shukla}, A.; {Mannheim}, K.; {Patel}, S.R.; {Roy}, J.; {Chitnis}, V.R.;
  {Dorner}, D.; {Rao}, A.R.; {Anupama}, G.C.; {Wendel}, C.
\newblock {Short-timescale {\ensuremath{\gamma}}-Ray Variability in CTA 102}.
\newblock {\em \apjl} {\bf 2018}, {\em 854},~L26.
\newblock
  doi:{\changeurlcolor{black}\href{https://doi.org/10.3847/2041-8213/aaacca}{\detokenize{10.3847/2041-8213/aaacca}}}.

\bibitem[{Marscher} and {Gear}(1985)]{1985ApJ...298..114M}
{Marscher}, A.P.; {Gear}, W.K.
\newblock {Models for high-frequency radio outbursts in extragalactic sources,
  with application to the early 1983 millimeter-to-infrared flare of 3C 273}.
\newblock {\em \apj} {\bf 1985}, {\em 298},~114--127.
\newblock
  doi:{\changeurlcolor{black}\href{https://doi.org/10.1086/163592}{\detokenize{10.1086/163592}}}.

\bibitem[{Giannios}(2013)]{2013MNRAS.431..355G}
{Giannios}, D.
\newblock {Reconnection-driven plasmoids in blazars: fast flares on a slow
  envelope}.
\newblock {\em \mnras} {\bf 2013}, {\em 431},~355--363.
{https://doi.org/10.1093/mnras/stt167}.

\bibitem[{Sironi} \em{et~al.}(2015){Sironi}, {Petropoulou}, and
  {Giannios}]{2015MNRAS.450..183S}
{Sironi}, L.; {Petropoulou}, M.; {Giannios}, D.
\newblock {Relativistic jets shine through shocks or magnetic reconnection?}
\newblock {\em \mnras} {\bf 2015}, {\em 450},~183--191.
{https://doi.org/10.1093/mnras/stv641}.

\bibitem[{Werner} \em{et~al.}(2016){Werner}, {Uzdensky}, {Cerutti},
  {Nalewajko}, and {Begelman}]{2016ApJ...816L...8W}
{Werner}, G.R.; {Uzdensky}, D.A.; {Cerutti}, B.; {Nalewajko}, K.; {Begelman},
  M.C.
\newblock {The Extent of Power-law Energy Spectra in Collisionless Relativistic
  Magnetic Reconnection in Pair Plasmas}.
\newblock {\em \apjl} {\bf 2016}, {\em 816},~L8.
{https://doi.org/10.3847/2041-8205/816/1/L8}.

\bibitem[{Shukla} and {Mannheim}(2020)]{2020NatCo..11.4176S}
{Shukla}, A.; {Mannheim}, K.
\newblock {Gamma-ray flares from relativistic magnetic reconnection in the jet
  of the quasar 3C 279}.
\newblock {\em Nature Communications} {\bf 2020}, {\em 11},~4176.
{https://doi.org/10.1038/s41467-020-17912-z}.

\bibitem[{Giannios} \em{et~al.}(2010){Giannios}, {Uzdensky}, and
  {Begelman}]{2010MNRAS.402.1649G}
{Giannios}, D.; {Uzdensky}, D.A.; {Begelman}, M.C.
\newblock {Fast TeV variability from misaligned minijets in the jet of M87}.
\newblock {\em \mnras} {\bf 2010}, {\em 402},~1649--1656.
{https://doi.org/10.1111/j.1365-2966.2009.16045.x}.

\bibitem[{Aharonian} \em{et~al.}(2007){Aharonian}, {Akhperjanian},
  {Bazer-Bachi}, {Behera}, {Beilicke}, {Benbow}, {Berge}, {Bernl{\"o}hr},
  {Boisson}, {Bolz}, {Borrel}, {Boutelier}, {Braun}, {Brion}, {Brown},
  {B{\"u}hler}, {B{\"u}sching}, {Bulik}, {Carrigan}, {Chadwick}, {Clapson},
  {Chounet}, {Coignet}, {Cornils}, {Costamante}, {Degrange}, {Dickinson},
  {Djannati-Ata{\"i}}, {Domainko}, {Drury}, {Dubus}, {Dyks}, {Egberts},
  {Emmanoulopoulos}, {Espigat}, {Farnier}, {Feinstein}, {Fiasson},
  {F{\"o}rster}, {Fontaine}, {Funk}, {Funk}, {F{\"u}{\ss}ling}, {Gallant},
  {Giebels}, {Glicenstein}, {Gl{\"u}ck}, {Goret}, {Hadjichristidis}, {Hauser},
  {Hauser}, {Heinzelmann}, {Henri}, {Hermann}, {Hinton}, {Hoffmann}, {Hofmann},
  {Holleran}, {Hoppe}, {Horns}, {Jacholkowska}, {de Jager}, {Kendziorra},
  {Kerschhaggl}, {Kh{\'e}lifi}, {Komin}, {Kosack}, {Lamanna}, {Latham}, {Le
  Gallou}, {Lemi{\`e}re}, {Lemoine-Goumard}, {Lenain}, {Lohse}, {Martin},
  {Martineau-Huynh}, {Marcowith}, {Masterson}, {Maurin}, {McComb}, {Moderski},
  {Moulin}, {de Naurois}, {Nedbal}, {Nolan}, {Olive}, {Orford}, {Osborne},
  {Ostrowski}, {Panter}, {Pedaletti}, {Pelletier}, {Petrucci}, {Pita},
  {P{\"u}hlhofer}, {Punch}, {Ranchon}, {Raubenheimer}, {Raue}, {Rayner},
  {Renaud}, {Ripken}, {Rob}, {Rolland}, {Rosier-Lees}, {Rowell}, {Rudak},
  {Ruppel}, {Sahakian}, {Santangelo}, {Saug{\'e}}, {Schlenker}, {Schlickeiser},
  {Schr{\"o}der}, {Schwanke}, {Schwarzburg}, {Schwemmer}, {Shalchi}, {Sol},
  {Spangler}, {Stawarz}, {Steenkamp}, {Stegmann}, {Superina}, {Tam},
  {Tavernet}, {Terrier}, {van Eldik}, {Vasileiadis}, {Venter}, {Vialle},
  {Vincent}, {Vivier}, {V{\"o}lk}, {Volpe}, {Wagner}, {Ward}, and
  {Zdziarski}]{2007ApJ...664L..71A}
{Aharonian}, F.; {Akhperjanian}, A.G.; {Bazer-Bachi}, A.R.; {Behera}, B.;
  {Beilicke}, M.; {Benbow}, W.; {Berge}, D.; {Bernl{\"o}hr}, K.; {Boisson}, C.;
  {Bolz}, O.;
  et~al.
\newblock {An Exceptional Very High Energy Gamma-Ray Flare of PKS 2155-304}.
\newblock {\em \apjl} {\bf 2007}, {\em 664},~L71--L74.
{https://doi.org/10.1086/520635}.

\bibitem[{Arlen} \em{et~al.}(2013){Arlen}, {Aune}, {Beilicke}, {Benbow},
  {Bouvier}, {Buckley}, {Bugaev}, {Cesarini}, {Ciupik}, {Connolly}, {Cui},
  {Dickherber}, {Dumm}, {Errando}, {Falcone}, {Federici}, {Feng}, {Finley},
  {Finnegan}, {Fortson}, {Furniss}, {Galante}, {Gall}, {Griffin}, {Grube},
  {Gyuk}, {Hanna}, {Holder}, {Humensky}, {Kaaret}, {Karlsson}, {Kertzman},
  {Khassen}, {Kieda}, {Krawczynski}, {Krennrich}, {Maier}, {Moriarty},
  {Mukherjee}, {Nelson}, {O'Faol{\'a}in de Bhr{\'o}ithe}, {Ong}, {Orr}, {Park},
  {Perkins}, {Pichel}, {Pohl}, {Prokoph}, {Quinn}, {Ragan}, {Reyes},
  {Reynolds}, {Roache}, {Saxon}, {Schroedter}, {Sembroski}, {Staszak},
  {Telezhinsky}, {Te{\v{s}}i{\'c}}, {Theiling}, {Tsurusaki}, {Varlotta},
  {Vincent}, {Wakely}, {Weekes}, {Weinstein}, {Welsing}, {Williams}, {Zitzer},
  {VERITAS Collaboration}, {Jorstad}, {MacDonald}, {Marscher}, {Smith},
  {Walker}, {Hovatta}, {Richards}, {Max-Moerbeck}, {Readhead}, {Lister},
  {Kovalev}, {Pushkarev}, {Gurwell}, {L{\"a}hteenm{\"a}ki}, {Nieppola},
  {Tornikoski}, and {J{\"a}rvel{\"a}}]{2013ApJ...762...92A}
{Arlen}, T.; {Aune}, T.; {Beilicke}, M.; {Benbow}, W.; {Bouvier}, A.;
  {Buckley}, J.H.; {Bugaev}, V.; {Cesarini}, A.; {Ciupik}, L.; {Connolly},
   et~al.
\newblock {Rapid TeV Gamma-Ray Flaring of BL Lacertae}.
\newblock {\em \apj} {\bf 2013}, {\em 762},~92.
{https://doi.org/10.1088/0004-637X/762/2/92}.

\bibitem[{Ackermann} \em{et~al.}(2016){Ackermann}, {Anantua}, {Asano},
  {Baldini}, {Barbiellini}, {Bastieri}, {Becerra Gonzalez}, {Bellazzini},
  {Bissaldi}, {Blandford}, {Bloom}, {Bonino}, {Bottacini}, {Bruel}, {Buehler},
  {Caliandro}, {Cameron}, {Caragiulo}, {Caraveo}, {Cavazzuti}, {Cecchi},
  {Cheung}, {Chiang}, {Chiaro}, {Ciprini}, {Cohen-Tanugi}, {Costanza},
  {Cutini}, {D'Ammando}, {de Palma}, {Desiante}, {Digel}, {Di Lalla}, {Di
  Mauro}, {Di Venere}, {Drell}, {Favuzzi}, {Fegan}, {Ferrara}, {Fukazawa},
  {Funk}, {Fusco}, {Gargano}, {Gasparrini}, {Giglietto}, {Giordano},
  {Giroletti}, {Grenier}, {Guillemot}, {Guiriec}, {Hayashida}, {Hays}, {Horan},
  {J{\'o}hannesson}, {Kensei}, {Kocevski}, {Kuss}, {La Mura}, {Larsson},
  {Latronico}, {Li}, {Longo}, {Loparco}, {Lott}, {Lovellette}, {Lubrano},
  {Madejski}, {Magill}, {Maldera}, {Manfreda}, {Mayer}, {Mazziotta},
  {Michelson}, {Mirabal}, {Mizuno}, {Monzani}, {Morselli}, {Moskalenko},
  {Nalewajko}, {Negro}, {Nuss}, {Ohsugi}, {Orlando}, {Paneque}, {Perkins},
  {Pesce-Rollins}, {Piron}, {Pivato}, {Porter}, {Principe}, {Rando}, {Razzano},
  {Razzaque}, {Reimer}, {Scargle}, {Sgr{\`o}}, {Sikora}, {Simone}, {Siskind},
  {Spada}, {Spinelli}, {Stawarz}, {Thayer}, {Thompson}, {Torres}, {Troja},
  {Uchiyama}, {Yuan}, and {Zimmer}]{2016ApJ...824L..20A}
{Ackermann}, M.; {Anantua}, R.; {Asano}, K.; {Baldini}, L.; {Barbiellini}, G.;
  {Bastieri}, D.; {Becerra Gonzalez}, J.; {Bellazzini}, R.; {Bissaldi}, E.;
  {Blandford}, R.D.;
  et~al.
\newblock {Minute-timescale >100 MeV {\ensuremath{\gamma}}-Ray Variability
  during the Giant Outburst of Quasar 3C 279 Observed by Fermi-LAT in 2015
  June}.
\newblock {\em \apjl} {\bf 2016}, {\em 824},~L20.
{https://doi.org/10.3847/2041-8205/824/2/L20}.

\bibitem[{Liu} and {Bai}(2006)]{2006ApJ...653.1089L}
{Liu}, H.T.; {Bai}, J.M.
\newblock {Absorption of 10-200 GeV Gamma Rays by Radiation from Broad-Line
  Regions in Blazars}.
\newblock {\em \apj} {\bf 2006}, {\em 653},~1089--1097.
{https://doi.org/10.1086/509097}.

\bibitem[{Abdollahi} \em{et~al.}(2020){Abdollahi}, {Acero}, {Ackermann},
  {Ajello}, {Atwood}, {Axelsson}, {Baldini}, {Ballet}, {Barbiellini},
  {Bastieri}, {Becerra Gonzalez}, {Bellazzini}, {Berretta}, {Bissaldi},
  {Blandford}, {Bloom}, {Bonino}, {Bottacini}, {Brandt}, {Bregeon}, {Bruel},
  {Buehler}, {Burnett}, {Buson}, {Cameron}, {Caputo}, {Caraveo}, {Casandjian},
  {Castro}, {Cavazzuti}, {Charles}, {Chaty}, {Chen}, {Cheung}, {Chiaro},
  {Ciprini}, {Cohen-Tanugi}, {Cominsky}, {Coronado-Bl{\'a}zquez}, {Costantin},
  {Cuoco}, {Cutini}, {D'Ammando}, {DeKlotz}, {de la Torre Luque}, {de Palma},
  {Desai}, {Digel}, {Di Lalla}, {Di Mauro}, {Di Venere}, {Dom{\'\i}nguez},
  {Dumora}, {Fana Dirirsa}, {Fegan}, {Ferrara}, {Franckowiak}, {Fukazawa},
  {Funk}, {Fusco}, {Gargano}, {Gasparrini}, {Giglietto}, {Giommi}, {Giordano},
  {Giroletti}, {Glanzman}, {Green}, {Grenier}, {Griffin}, {Grondin}, {Grove},
  {Guiriec}, {Harding}, {Hayashi}, {Hays}, {Hewitt}, {Horan},
  {J{\'o}hannesson}, {Johnson}, {Kamae}, {Kerr}, {Kocevski}, {Kovac'evic'},
  {Kuss}, {Landriu}, {Larsson}, {Latronico}, {Lemoine-Goumard}, {Li},
  {Liodakis}, {Longo}, {Loparco}, {Lott}, {Lovellette}, {Lubrano}, {Madejski},
  {Maldera}, {Malyshev}, {Manfreda}, {Marchesini}, {Marcotulli},
  {Mart{\'\i}-Devesa}, {Martin}, {Massaro}, {Mazziotta}, {McEnery}, {Mereu},
  {Meyer}, {Michelson}, {Mirabal}, {Mizuno}, {Monzani}, {Morselli},
  {Moskalenko}, {Negro}, {Nuss}, {Ojha}, {Omodei}, {Orienti}, {Orlando},
  {Ormes}, {Palatiello}, {Paliya}, {Paneque}, {Pei}, {Pe{\~n}a-Herazo},
  {Perkins}, {Persic}, {Pesce-Rollins}, {Petrosian}, {Petrov}, {Piron}, {Poon},
  {Porter}, {Principe}, {Rain{\`o}}, {Rando}, {Razzano}, {Razzaque}, {Reimer},
  {Reimer}, {Remy}, {Reposeur}, {Romani}, {Saz Parkinson}, {Schinzel},
  {Serini}, {Sgr{\`o}}, {Siskind}, {Smith}, {Spandre}, {Spinelli}, {Strong},
  {Suson}, {Tajima}, {Takahashi}, {Tak}, {Thayer}, {Thompson}, {Tibaldo},
  {Torres}, {Torresi}, {Valverde}, {Van Klaveren}, {van Zyl}, {Wood},
  {Yassine}, and {Zaharijas}]{2020ApJS..247...33A}
{Abdollahi}, S.; {Acero}, F.; {Ackermann}, M.; {Ajello}, M.; {Atwood}, W.B.;
  {Axelsson}, M.; {Baldini}, L.; {Ballet}, J.; {Barbiellini}, G.; {Bastieri},
  D.;
  et~al.
\newblock {Fermi Large Area Telescope Fourth Source Catalog}.
\newblock {\em \apjs} {\bf 2020}, {\em 247},~33.
{https://doi.org/10.3847/1538-4365/ab6bcb}.

\bibitem[{Hewett} and {Wild}(2010)]{2010MNRAS.405.2302H}
{Hewett}, P.C.; {Wild}, V.
\newblock {Improved redshifts for SDSS quasar spectra}.
\newblock {\em \mnras} {\bf 2010}, {\em 405},~2302--2316.
{https://doi.org/10.1111/j.1365-2966.2010.16648.x}.

\bibitem[{Fan} \em{et~al.}(2006){Fan}, {Tao}, {Qian}, {Gupta}, {Liu}, {Yuan},
  {Yang}, {Wang}, and {Huang}]{2006PASJ...58..797F}
{Fan}, J.H.; {Tao}, J.; {Qian}, B.C.; {Gupta}, A.C.; {Liu}, Y.; {Yuan}, Y.H.;
  {Yang}, J.H.; {Wang}, H.G.; {Huang}, Y.
\newblock {Optical Photometrical Observations and Variability for Quasar 4C
  29.45}.
\newblock {\em Publ. Astron. Soc. Jpn.} {\bf 2006}, {\em 58},~797--808.
{https://doi.org/10.1093/pasj/58.5.797}.

\bibitem[{Thompson} \em{et~al.}(1995){Thompson}, {Bertsch}, {Dingus},
  {Esposito}, {Etienne}, {Fichtel}, {Friedlander}, {Hartman}, {Hunter},
  {Kendig}, {Mattox}, {McDonald}, {von Montigny}, {Mukherjee}, {Ramanamurthy},
  {Sreekumar}, {Fierro}, {Lin}, {Michelson}, {Nolan}, {Shriver}, {Willis},
  {Kanbach}, {Mayer-Hasselwander}, {Merck}, {Radecke}, {Kniffen}, and
  {Schneid}]{1995ApJS..101..259T}
{Thompson}, D.J.; {Bertsch}, D.L.; {Dingus}, B.L.; {Esposito}, J.A.; {Etienne},
  A.; {Fichtel}, C.E.; {Friedlander}, D.P.; {Hartman}, R.C.; {Hunter}, S.D.;
  {Kendig}, D.J.;
  et~al.
\newblock {The Second EGRET Catalog of High-Energy Gamma-Ray Sources}.
\newblock {\em \apjs} {\bf 1995}, {\em 101},~259.
{https://doi.org/10.1086/192240}.

\bibitem[{Abdo} \em{et~al.}(2010){Abdo}, {Ackermann}, {Ajello}, {Allafort},
  {Antolini}, {Atwood}, {Axelsson}, {Baldini}, {Ballet}, {Barbiellini},
  {Bastieri}, {Baughman}, {Bechtol}, {Bellazzini}, {Berenji}, {Blandford},
  {Bloom}, {Bogart}, {Bonamente}, {Borgland}, {Bouvier}, {Bregeon}, {Brez},
  {Brigida}, {Bruel}, {Buehler}, {Burnett}, {Buson}, {Caliandro}, {Cameron},
  {Cannon}, {Caraveo}, {Carrigan}, {Casandjian}, {Cavazzuti}, {Cecchi}, {{\c
  C}elik}, {Celotti}, {Charles}, {Chekhtman}, {Chen}, {Cheung}, {Chiang},
  {Ciprini}, {Claus}, {Cohen-Tanugi}, {Conrad}, {Costamante}, {Cotter},
  {Cutini}, {D'Elia}, {Dermer}, {de Angelis}, {de Palma}, {De Rosa}, {Digel},
  {Silva}, {Drell}, {Dubois}, {Dumora}, {Escande}, {Farnier}, {Favuzzi},
  {Fegan}, {Ferrara}, {Focke}, {Fortin}, {Frailis}, {Fukazawa}, {Funk},
  {Fusco}, {Gargano}, {Gasparrini}, {Gehrels}, {Germani}, {Giebels},
  {Giglietto}, {Giommi}, {Giordano}, {Giroletti}, {Glanzman}, {Godfrey},
  {Grandi}, {Grenier}, {Grondin}, {Grove}, {Guiriec}, {Hadasch}, {Harding},
  {Hayashida}, {Hays}, {Healey}, {Hill}, {Horan}, {Hughes}, {Iafrate}, {Itoh},
  {J{\'o}hannesson}, {Johnson}, {Johnson}, {Johnson}, {Johnson}, {Kamae},
  {Katagiri}, {Kataoka}, {Kawai}, {Kerr}, {Kn{\"o}dlseder}, {Kuss}, {Lande},
  {Latronico}, {Lavalley}, {Lemoine-Goumard}, {Llena Garde}, {Longo},
  {Loparco}, {Lott}, {Lovellette}, {Lubrano}, {Madejski}, {Makeev}, {Malaguti},
  {Massaro}, {Mazziotta}, {McConville}, {McEnery}, {McGlynn}, {Michelson},
  {Mitthumsiri}, {Mizuno}, {Moiseev}, {Monte}, {Monzani}, {Morselli},
  {Moskalenko}, {Murgia}, {Nolan}, {Norris}, {Nuss}, {Ohno}, {Ohsugi},
  {Omodei}, {Orlando}, {Ormes}, {Ozaki}, {Paneque}, {Panetta}, {Parent},
  {Pelassa}, {Pepe}, {Pesce-Rollins}, {Piranomonte}, {Piron}, {Porter},
  {Rain{\`o}}, {Rando}, {Razzano}, {Reimer}, {Reimer}, {Reposeur}, {Ripken},
  {Ritz}, {Rodriguez}, {Romani}, {Roth}, {Ryde}, {Sadrozinski}, {Sanchez},
  {Sander}, {Saz Parkinson}, {Scargle}, {Sgr{\`o}}, {Shaw}, {Siskind}, {Smith},
  {Spandre}, {Spinelli}, {Starck}, {Stawarz}, {Strickman}, {Suson}, {Tajima},
  {Takahashi}, {Takahashi}, {Tanaka}, {Taylor}, {Thayer}, {Thayer}, {Thompson},
  {Tibaldo}, {Torres}, {Tosti}, {Tramacere}, {Ubertini}, {Uchiyama}, {Usher},
  {Vasileiou}, {Vilchez}, {Villata}, {Vitale}, {Waite}, {Wallace}, {Wang},
  {Winer}, {Wood}, {Yang}, {Ylinen}, and {Ziegler}]{2010ApJ...715..429A}
{Abdo}, A.A.; {Ackermann}, M.; {Ajello}, M.; {Allafort}, A.; {Antolini}, E.;
  {Atwood}, W.B.; {Axelsson}, M.; {Baldini}, L.; {Ballet}, J.; {Barbiellini},
  G.;
  et~al.
\newblock {The First Catalog of Active Galactic Nuclei Detected by the Fermi
  Large Area Telescope}.
\newblock {\em \apj} {\bf 2010}, {\em 715},~429--457.
{https://doi.org/10.1088/0004-637X/715/1/429}.

\bibitem[{Ramakrishnan} \em{et~al.}(2014){Ramakrishnan}, {Le{\'o}n-Tavares},
  {Rastorgueva-Foi}, {Wiik}, {Jorstad}, {Marscher}, {Tornikoski}, {Agudo},
  {L{\"a}hteenm{\"a}ki}, {Valtaoja}, {Aller}, {Blinov}, {Casadio}, {Efimova},
  {Gurwell}, {G{\'o}mez}, {Hagen-Thorn}, {Joshi}, {J{\"a}rvel{\"a}},
  {Konstantinova}, {Kopatskaya}, {Larionov}, {Larionova}, {Larionova},
  {Lavonen}, {MacDonald}, {McHardy}, {Molina}, {Morozova}, {Nieppola}, {Tammi},
  {Taylor}, and {Troitsky}]{2014MNRAS.445.1636R}
{Ramakrishnan}, V.; {Le{\'o}n-Tavares}, J.; {Rastorgueva-Foi}, E.A.; {Wiik},
  K.; {Jorstad}, S.G.; {Marscher}, A.P.; {Tornikoski}, M.; {Agudo}, I.;
  {L{\"a}hteenm{\"a}ki}, A.; {Valtaoja}, E.;
  et~al.
\newblock {The connection between the parsec-scale radio jet and
  {\ensuremath{\gamma}}-ray flares in the blazar 1156+295}.
\newblock {\em \mnras} {\bf 2014}, {\em 445},~1636--1646.
{https://doi.org/10.1093/mnras/stu1873}.

\bibitem[{Prince}(2019)]{2019ApJ...871..101P}
{Prince}, R.
\newblock {Multi-frequency Variability Study of Ton 599 during the High
  Activity of 2017}.
\newblock {\em \apj} {\bf 2019}, {\em 871},~101.
{https://doi.org/10.3847/1538-4357/aaf475}.

\bibitem[{Patel} \em{et~al.}(2018){Patel}, {Chitnis}, {Shukla}, {Rao}, and
  {Nagare}]{2018ApJ...866..102P}
{Patel}, S.R.; {Chitnis}, V.R.; {Shukla}, A.; {Rao}, A.R.; {Nagare}, B.J.
\newblock {Temporal Variability and Estimation of Jet Parameters for Ton 599}.
\newblock {\em \apj} {\bf 2018}, {\em 866},~102.
{https://doi.org/10.3847/1538-4357/aae1fc}.

\bibitem[{Patel} and {Chitnis}(2020)]{2020MNRAS.492...72P}
{Patel}, S.R.; {Chitnis}, V.R.
\newblock {Leptonic modelling of Ton 599 in flare and quiescent states}.
\newblock {\em \mnras} {\bf 2020}, {\em 492},~72--78.
\newblock
  doi:{\changeurlcolor{black}\href{https://doi.org/10.1093/mnras/stz3490}{\detokenize{10.1093/mnras/stz3490}}}.

\bibitem[{Atwood} \em{et~al.}(2009){Atwood}, {Abdo}, {Ackermann}, {Althouse},
  {Anderson}, {Axelsson}, {Baldini}, {Ballet}, {Band}, {Barbiellini}, and
  et~al.]{2009ApJ...697.1071A}
{Atwood}, W.B.; {Abdo}, A.A.; {Ackermann}, M.; {Althouse}, W.; {Anderson}, B.;
  {Axelsson}, M.; {Baldini}, L.; {Ballet}, J.; {Band}, D.L.; {Barbiellini}, G.;
  et~al.
\newblock {The Large Area Telescope on the Fermi Gamma-Ray Space Telescope
  Mission}.
\newblock {\em \apj} {\bf 2009}, {\em 697},~1071--1102.
{https://doi.org/10.1088/0004-637X/697/2/1071}.

\bibitem[{Mattox} \em{et~al.}(1996){Mattox}, {Bertsch}, {Chiang}, {Dingus},
  {Digel}, {Esposito}, {Fierro}, {Hartman}, {Hunter}, {Kanbach}, {Kniffen},
  {Lin}, {Macomb}, {Mayer-Hasselwander}, {Michelson}, {von Montigny},
  {Mukherjee}, {Nolan}, {Ramanamurthy}, {Schneid}, {Sreekumar}, {Thompson}, and
  {Willis}]{1996ApJ...461..396M}
{Mattox}, J.R.; {Bertsch}, D.L.; {Chiang}, J.; {Dingus}, B.L.; {Digel}, S.W.;
  {Esposito}, J.A.; {Fierro}, J.M.; {Hartman}, R.C.; {Hunter}, S.D.; {Kanbach},
  G.;
  et~al.
\newblock {The Likelihood Analysis of EGRET Data}.
\newblock {\em \apj} {\bf 1996}, {\em 461},~396.
{https://doi.org/10.1086/177068}.

\bibitem[{Vaughan} \em{et~al.}(2003){Vaughan}, {Edelson}, {Warwick}, and
  {Uttley}]{2003MNRAS.345.1271V}
{Vaughan}, S.; {Edelson}, R.; {Warwick}, R.S.; {Uttley}, P.
\newblock {On characterizing the variability properties of X-ray light curves
  from active galaxies}.
\newblock {\em \mnras} {\bf 2003}, {\em 345},~1271--1284.
{https://doi.org/10.1046/j.1365-2966.2003.07042.x}.

\bibitem[{Pandey} \em{et~al.}(2018){Pandey}, {Gupta}, and
  {Wiita}]{2018ApJ...859...49P}
{Pandey}, A.; {Gupta}, A.C.; {Wiita}, P.J.
\newblock {X-Ray Flux and Spectral Variability of Six TeV Blazars with NuSTAR}.
\newblock {\em \apj} {\bf 2018}, {\em 859},~49.
{https://doi.org/10.3847/1538-4357/aabc5b}.

\bibitem[{Nolan} \em{et~al.}(2012){Nolan}, {Abdo}, {Ackermann}, {Ajello},
  {Allafort}, {Antolini}, {Atwood}, {Axelsson}, {Baldini}, {Ballet}, and
  et~al.]{2012ApJS..199...31N}
{Nolan}, P.L.; {Abdo}, A.A.; {Ackermann}, M.; {Ajello}, M.; {Allafort}, A.;
  {Antolini}, E.; {Atwood}, W.B.; {Axelsson}, M.; {Baldini}, L.; {Ballet}, J.;
  et~al.
\newblock {Fermi Large Area Telescope Second Source Catalog}.
\newblock {\em \apjs} {\bf 2012}, {\em 199},~31.
https://doi.org/10.1088/0067-0049/199/2/31.

\bibitem[{Paliya}(2015)]{2015ApJ...808L..48P}
{Paliya}, V.S.
\newblock {Fermi-Large Area Telescope Observations of the Exceptional Gamma-Ray
  Flare from 3C 279 in 2015 June}.
\newblock {\em \apjl} {\bf 2015}, {\em 808},~L48.
{https://doi.org/10.1088/2041-8205/808/2/L48}.

\bibitem[{Zhang} \em{et~al.}(2002){Zhang}, {Fan}, and
  {Cheng}]{2002PASJ...54..159Z}
{Zhang}, L.Z.; {Fan}, J.H.; {Cheng}, K.S.
\newblock {The Multiwavelength Doppler Factors for a Sample of Gamma-Ray Loud
  Blazars}.
\newblock {\em \pasj} {\bf 2002}, {\em 54},~159--169.
{https://doi.org/10.1093/pasj/54.2.159}.

\bibitem[{Foschini} \em{et~al.}(2011){Foschini}, {Ghisellini}, {Tavecchio},
  {Bonnoli}, and {Stamerra}]{2011A&A...530A..77F}
{Foschini}, L.; {Ghisellini}, G.; {Tavecchio}, F.; {Bonnoli}, G.; {Stamerra},
  A.
\newblock {Search for the shortest variability at gamma rays in flat-spectrum
  radio quasars}.
\newblock {\em \aap} {\bf 2011}, {\em 530},~A77.
{https://doi.org/10.1051/0004-6361/201117064}.

\bibitem[{Pushkarev} \em{et~al.}(2009){Pushkarev}, {Kovalev}, {Lister}, and
  {Savolainen}]{2009A&A...507L..33P}
{Pushkarev}, A.B.; {Kovalev}, Y.Y.; {Lister}, M.L.; {Savolainen}, T.
\newblock {Jet opening angles and gamma-ray brightness of AGN}.
\newblock {\em \aap} {\bf 2009}, {\em 507},~L33--L36. {https://doi.org/10.1051/0004-6361/200913422}.

\bibitem[{Ghisellini} and {Maraschi}(1989)]{1989ApJ...340..181G}
{Ghisellini}, G.; {Maraschi}, L.
\newblock {Bulk acceleration in relativistic jets and the spectral properties
  of blazars}.
\newblock {\em \apj} {\bf 1989}, {\em 340},~181--189.
{https://doi.org/10.1086/167383}.

\bibitem[{Begelman} \em{et~al.}(1987){Begelman}, {Sikora}, {Giommi}, {Barr},
  {Garilli}, {Gioia}, {Maccacaro}, {Maccagni}, and
  {Schild}]{1987ApJ...322..650B}
{Begelman}, M.C.; {Sikora}, M.; {Giommi}, P.; {Barr}, P.; {Garilli}, B.;
  {Gioia}, I.M.; {Maccacaro}, T.; {Maccagni}, D.; {Schild}, R.E.
\newblock {Inverse Compton scattering of ambient radiation by a cold
  relativistic jet - A source of beamed, polarized continuum in blazars?}
\newblock {\em \apj} {\bf 1987}, {\em 322},~650--661. {https://doi.org/10.1086/165760}.

\bibitem[{Sikora} \em{et~al.}(1994){Sikora}, {Begelman}, and
  {Rees}]{1994ApJ...421..153S}
{Sikora}, M.; {Begelman}, M.C.; {Rees}, M.J.
\newblock {Comptonization of diffuse ambient radiation by a relativistic jet:
  The source of gamma rays from blazars?}
\newblock {\em \apj} {\bf 1994}, {\em 421},~153--162.
\newblock
  doi:{\changeurlcolor{black}\href{https://doi.org/10.1086/173633}{\detokenize{10.1086/173633}}}.

\bibitem[{Sbarrato} \em{et~al.}(2011){Sbarrato}, {Foschini}, {Ghisellini}, and
  {Tavecchio}]{2011AdSpR..48..998S}
{Sbarrato}, T.; {Foschini}, L.; {Ghisellini}, G.; {Tavecchio}, F.
\newblock {Study of the variability of blazars gamma-ray emission}.
\newblock {\em Advances in Space Research} {\bf 2011}, {\em 48},~998--1003.
{https://doi.org/10.1016/j.asr.2011.05.019}.

\bibitem[{Abdo} \em{et~al.}(2010){Abdo}, {Ackermann}, {Ajello}, {Atwood},
  {Axelsson}, {Baldini}, {Ballet}, {Barbiellini}, {Bastieri}, {Bechtol},
  {Bellazzini}, {Berenji}, {Blandford}, {Bloom}, {Bonamente}, {Borgland},
  {Bouvier}, {Bregeon}, {Brez}, {Brigida}, {Bruel}, {Burnett}, {Buson},
  {Caliandro}, {Cameron}, {Caraveo}, {Carrigan}, {Casandjian}, {Cavazzuti},
  {Cecchi}, {{\c{C}}elik}, {Charles}, {Chekhtman}, {Cheung}, {Chiang},
  {Ciprini}, {Claus}, {Cohen-Tanugi}, {Conrad}, {Cutini}, {Dermer}, {de
  Angelis}, {de Palma}, {Digel}, {Silva}, {Drell}, {Dubois}, {Dumora},
  {Farnier}, {Favuzzi}, {Fegan}, {Focke}, {Fortin}, {Frailis}, {Fukazawa},
  {Funk}, {Fusco}, {Gargano}, {Gasparrini}, {Gehrels}, {Germani}, {Giebels},
  {Giglietto}, {Giommi}, {Giordano}, {Glanzman}, {Godfrey}, {Grenier},
  {Grondin}, {Grove}, {Guillemot}, {Guiriec}, {Harding}, {Hartman},
  {Hayashida}, {Hays}, {Healey}, {Horan}, {Hughes}, {Jackson},
  {J{\'o}hannesson}, {Johnson}, {Johnson}, {Kamae}, {Katagiri}, {Kataoka},
  {Kawai}, {Kerr}, {Kn{\"o}dlseder}, {Kuss}, {Lande}, {Latronico},
  {Lemoine-Goumard}, {Longo}, {Loparco}, {Lott}, {Lovellette}, {Lubrano},
  {Madejski}, {Makeev}, {Mazziotta}, {McConville}, {McEnery}, {Meurer},
  {Michelson}, {Mitthumsiri}, {Mizuno}, {Moiseev}, {Monte}, {Monzani},
  {Morselli}, {Moskalenko}, {Murgia}, {Nolan}, {Norris}, {Nuss}, {Ohsugi},
  {Omodei}, {Orlando}, {Ormes}, {Paneque}, {Panetta}, {Parent}, {Pelassa},
  {Pepe}, {Persic}, {Pesce-Rollins}, {Piron}, {Porter}, {Rain{\`o}}, {Rando},
  {Razzano}, {Reimer}, {Reimer}, {Reposeur}, {Ritz}, {Rochester}, {Rodriguez},
  {Romani}, {Roth}, {Ryde}, {Sadrozinski}, {Sanchez}, {Sander}, {Saz
  Parkinson}, {Scargle}, {Sgr{\`o}}, {Siskind}, {Smith}, {Smith}, {Spandre},
  {Spinelli}, {Strickman}, {Suson}, {Tajima}, {Takahashi}, {Takahashi},
  {Tanaka}, {Thayer}, {Thayer}, {Thompson}, {Tibaldo}, {Torres}, {Tosti},
  {Tramacere}, {Uchiyama}, {Usher}, {Vasileiou}, {Vilchez}, {Villata},
  {Vitale}, {Waite}, {Wang}, {Winer}, {Wood}, {Ylinen}, and
  {Ziegler}]{2010ApJ...710.1271A}
{Abdo}, A.A.; {Ackermann}, M.; {Ajello}, M.; {Atwood}, W.B.; {Axelsson}, M.;
  {Baldini}, L.; {Ballet}, J.; {Barbiellini}, G.; {Bastieri}, D.; {Bechtol},
  K.;
  et~al.
\newblock {Spectral Properties of Bright Fermi-Detected Blazars in the
  Gamma-Ray Band}.
\newblock {\em \apj} {\bf 2010}, {\em 710},~1271--1285.
{https://doi.org/10.1088/0004-637X/710/2/1271}.

\bibitem[{Paliya} \em{et~al.}(2015){Paliya}, {Sahayanathan}, and
  {Stalin}]{2015ApJ...803...15P}
{Paliya}, V.S.; {Sahayanathan}, S.; {Stalin}, C.S.
\newblock {Multi-Wavelength Observations of 3C 279 During the Extremely Bright
  Gamma-Ray Flare in 2014 March-April}.
\newblock {\em \apj} {\bf 2015}, {\em 803},~15.
{https://doi.org/10.1088/0004-637X/803/1/15}.

\bibitem[{Rajput} \em{et~al.}(2019){Rajput}, {Stalin}, {Sahayanathan},
  {Rakshit}, and {Mandal}]{2019MNRAS.486.1781R}
{Rajput}, B.; {Stalin}, C.S.; {Sahayanathan}, S.; {Rakshit}, S.; {Mandal}, A.K.
\newblock {Temporal correlation between the optical and
  {\ensuremath{\gamma}}-ray flux variations in the blazar 3C 454.3}.
\newblock {\em \mnras} {\bf 2019}, {\em 486},~1781--1795.
{https://doi.org/10.1093/mnras/stz941}.

\bibitem[{Coogan} \em{et~al.}(2016){Coogan}, {Brown}, and
  {Chadwick}]{2016MNRAS.458..354C}
{Coogan}, R.T.; {Brown}, A.M.; {Chadwick}, P.M.
\newblock {Localizing the {$\gamma$}-ray emission region during the 2014 June
  outburst of 3C 454.3}.
\newblock {\em \mnras} {\bf 2016}, {\em 458},~354--365.
{https://doi.org/10.1093/mnras/stw199}.

\bibitem[{Cerruti} \em{et~al.}(2013){Cerruti}, {Dermer}, {Lott}, {Boisson}, and
  {Zech}]{2013ApJ...771L...4C}
{Cerruti}, M.; {Dermer}, C.D.; {Lott}, B.; {Boisson}, C.; {Zech}, A.
\newblock {Gamma-Ray Blazars near Equipartition and the Origin of the GeV
  Spectral Break in 3C 454.3}.
\newblock {\em \apjl} {\bf 2013}, {\em 771},~L4. {https://doi.org/10.1088/2041-8205/771/1/L4}.

\bibitem[{Costamante} \em{et~al.}(2018){Costamante}, {Cutini}, {Tosti},
  {Antolini}, and {Tramacere}]{2018MNRAS.477.4749C}
{Costamante}, L.; {Cutini}, S.; {Tosti}, G.; {Antolini}, E.; {Tramacere}, A.
\newblock {On the origin of gamma-rays in Fermi blazars: beyondthe broad-line
  region}.
\newblock {\em \mnras} {\bf 2018}, {\em 477},~4749--4767. {https://doi.org/10.1093/mnras/sty887}.

\bibitem[{Tramacere} \em{et~al.}(2009){Tramacere}, {Giommi}, {Perri},
  {Verrecchia}, and {Tosti}]{2009A&A...501..879T}
{Tramacere}, A.; {Giommi}, P.; {Perri}, M.; {Verrecchia}, F.; {Tosti}, G.
\newblock {Swift observations of the very intense flaring activity of Mrk 421
  during 2006. I. Phenomenological picture of electron acceleration and
  predictions for MeV/GeV emission}.
\newblock {\em \aap} {\bf 2009}, {\em 501},~879--898.
{https://doi.org/10.1051/0004-6361/200810865}.

\bibitem[{Dermer} \em{et~al.}(2015){Dermer}, {Yan}, {Zhang}, {Finke}, and
  {Lott}]{2015ApJ...809..174D}
{Dermer}, C.D.; {Yan}, D.; {Zhang}, L.; {Finke}, J.D.; {Lott}, B.
\newblock {Near-equipartition Jets with Log-parabola Electron Energy
  Distribution and the Blazar Spectral-index Diagrams}.
\newblock {\em \apj} {\bf 2015}, {\em 809},~174.
{https://doi.org/10.1088/0004-637X/809/2/174}.

\end{thebibliography}
\end{document}